\documentclass{article}

\usepackage{arxiv}

\usepackage[utf8]{inputenc} 
\usepackage[T1]{fontenc}    
\usepackage{hyperref}       
\usepackage{url}            
\usepackage{booktabs}       
\usepackage{amsfonts}       
\usepackage{nicefrac}       
\usepackage{microtype}      
\usepackage{lipsum}
\usepackage{amsmath}
\usepackage{graphicx}
\usepackage{fixltx2e}
\usepackage{caption}

\graphicspath{ {./images/} }

\title{Differential Shape Optimization with Image Representation for Photonic Design }

\author{
Zhaocheng Liu \\
  Meta Reality Labs\\
  Redmond, WA, 98052\\
  \texttt{zhaocheng@meta.com} \\
   \And
Jim Bonar \\
  Meta Reality Labs\\
  Redmond, WA, 98052\\
  \texttt{jimbonar@meta.com} \\
}

\begin{document}
\maketitle

\begin{abstract}
We propose a general framework for differentiating shapes represented in binary images with respect to their parameters. This framework functions as an automatic differentiation tool for shape parameters, generating both binary density maps for optical simulations and computing gradients when the simulation provides a gradient of the density map. Our algorithm enables robust gradient computation that is insensitive to the image's pixel resolution and is compatible with all density-based simulation methods. We demonstrate the accuracy, effectiveness, and generalizability of our differential shape algorithm using photonic designs with different shape parametrizations across several differentiable optical solvers. We also demonstrate a substantial reduction in optimization time using our gradient-based shape optimization framework compared to traditional black-box optimization methods.
\end{abstract}


\section{Introduction}
In modern photonic design, structures are becoming increasingly complex, and performance requirements are more demanding. As a result, optical and photonic systems require a growing number of design parameters. While traditional black-box optimization techniques can effectively identify global optima for a small set of parameters, they become computationally intensive as the number of parameters increases \cite{frazier2018tutorial}, making it challenging to find optimal solutions within reasonable time and cost constraints. As such, gradient-based optimization is the preferred method for optimizing large numbers of parameters, as seen in neural networks used in modern applications such as human intelligence \cite{silver2016mastering} and generative AI \cite{touvron2023llama}\cite{vaswani2017attention}\cite{rombach2022high}. In these cases, millions or even billions of parameters need to be optimized, making gradient-based optimization a more efficient and effective approach \cite{kingma2014adam}.

To enhance optimization in physical systems, recent developments have also focused on gradient-based optimization methods for photonics design. A key advancement in this area is inverse design and topology optimization \cite{molesky2018inverse}\cite{christiansen2021inverse}, where the design is parameterized by freeform pixel images, and the topology is optimized using physical gradients calculated from the adjoint method or similar approaches. Topology optimization offers the maximum degree of freedom in design and often results in substantial performance enhancements \cite{piggott2018inverse}\cite{hammondhigh2022}\cite{pestourie2021inverse}\cite{phan2019high}\cite{sell2017large}. Concurrently, machine learning-based optimization is emerging as another promising area where gradient-based techniques are applied to discover complex photonic structures\cite{ma2021deep}\cite{wiecha2021deep}\cite{jiang2021deep}\cite{li2022empowering}\cite{zhelyeznyakov2023large}\cite{sanchez2024advances}. Some methods leverage the rapid gradient computation of neural surrogate models to provide gradient information for optimization\cite{an2019deep}\cite{liu2018training}\cite{malkiel2018plasmonic}, while others parameterize designs using neural networks to achieve rich representations of photonic structures \cite{ma2019probabilistic}\cite{liu2018generative}\cite{jiang2019global}. Gradient-based design approaches, including both machine learning-based optimization and inverse design, demonstrates significant potential for achieving high performant, demanding designs for applications in imaging \cite{tseng2021differentiable}\cite{yang2024curriculum}, sensing \cite{arya2024large}, perception \cite{chen2023meta}, display \cite{li2022inverse}\cite{zhu2021building}, and analog/quantum computing system \cite{chakravarthi2020inverse}\cite{nikkhah2024inverse}. 

In the meantime, one of the key challenges in gradient-based design is the lack of a general framework for calculating gradients with respect to shape parameters. Many optical and photonic designs are parametrized by shapes, which can involve numerous optimizable parameters that are difficult to handle with non-gradient-based optimization methods. Several approaches exist for deriving shape gradients, such as binarizing a level set function \cite{lalau2013adjoint}, deriving gradients for specific shapes \cite{mansouree2021large}, parametrizing shapes using control points \cite{luce2023merging}, or applying finite difference methods \cite{michaels2018leveraging}. While these techniques can yield accurate gradients, they often require tuning hyperparameters, are limited to specific geometries, or demand significant computational resources for large design spaces.

To address these challenges, this paper introduces a general framework for differential optimization of shape parameters using image representations, applicable to both explicit and implicit shape functions, and compatible with all density-based simulation methods. We begin by detailing the process for forming image representations and calculating gradients of shape parameters for explicit shape representations. Next, we demonstrate the efficacy and versatility of our framework through various design examples using different differential solvers, including FDFD, FDTD, and RCWA. Finally, we extend our algorithm to calculate gradients for shapes represented by implicit functions, enabling more general shape optimization.

\section{Differential Shape Optimization}
\label{sec:headings}

\subsection{Overall Workflow }
The goal of differential shape optimization is to efficiently calculate the gradients of shape parameters for optical and photonic design. Figure 1 illustrates the forward and backward workflow of this process. In the forward computation, we first define the optimizable parameters and then render the shapes on a canvas represented by an image. This image is subsequently fed into a simulation method for modeling. If the simulation is differentiable, the gradient can be derived using adjoint method or automatic differentiation. The resulting gradient map, which mirrors the shape of the forward-rendered image, is then back propagated to the shape parameters

\begin{figure}
  \label{fig:1}
  \includegraphics[width=16cm]{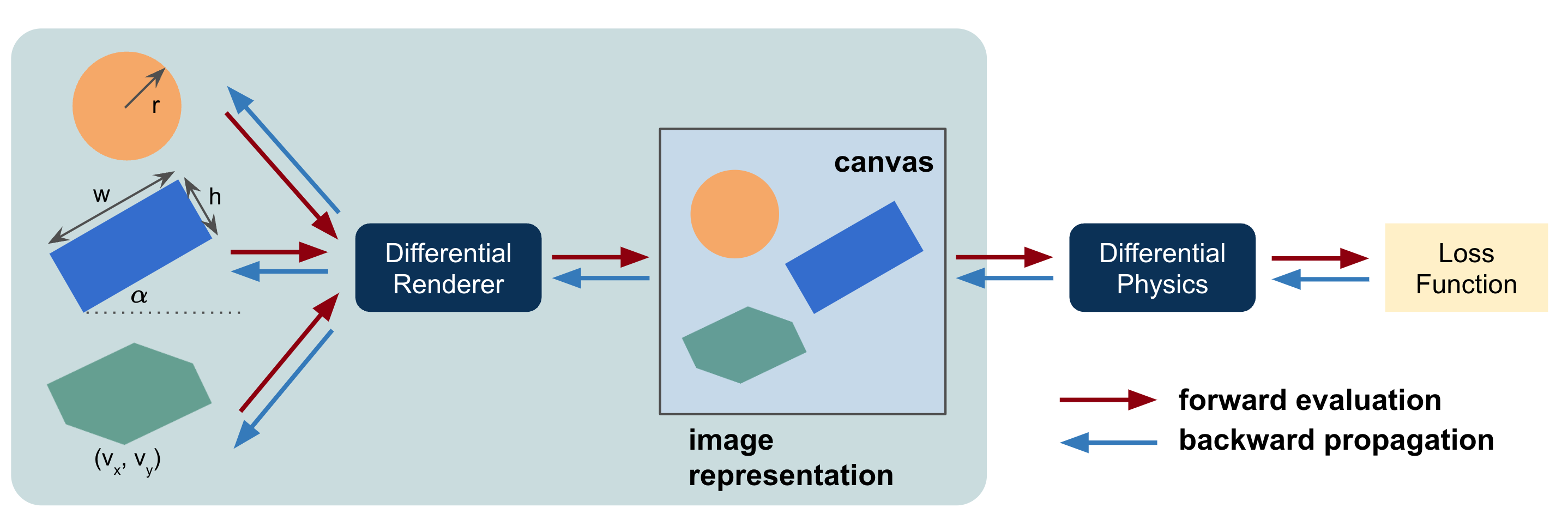}
  \caption{ Overview of the differential shape optimization pipeline. Our algorithmic framework enables the generation of photonic structures in image representations and supports gradient computation for all shape parameters, provided the simulation is differentiable.}
\end{figure}

Despite the simplicity of the concept, generating gradients for shape parameters is quite challenging. Two major obstacles contribute to this difficulty: (1) the image representation of a shape is inherently binary, with hard edges that impede gradient propagation, and (2) the gradient of a shape parameter follows a shape derivative, which is not directly compatible with the chain rule. In this paper, we introduce several techniques and algorithms to address these challenges.

\subsection{Shape Derivatives}
Before delving into the full algorithm for computing shape gradients in image representations, we briefly introduce shape derivatives prior to rendering the shape onto an image. We chose to present the derivative using Raynold’s transport theorem.

The core of the shape derivative is to determine the gradient of a parameter with respect to perturbations in the shape. Consider a function $f$ bounded over the domain $\Omega(x)$, with a parameter $p$, and an objective function function involving the integration of $f$:

\begin{equation} \label{eq1}
    L = \int_\Omega f(x, p) dV   
\end{equation}

\noindent where $V$ represents the volume or area of the interior of $\Omega$ in 3D and 2D scenarios, respectively. This formulation encompasses the definition of many optimization objective functions in optical and photonic designs. According to Raynold's theorem, the gradient of the loss function with respect to the parameters can be expressed as:

\begin{equation} \label{eq2}
    \frac{d}{dp}L = \int_\Omega \frac{\partial}{\partial p} f(x, p)dV  + \int_{\partial \Omega} (v_b \cdot n) f(x, p) dA
\end{equation}
where $\partial \Omega$ denotes the boundary of the domain, $n$ represents the normal direction of the surface, and $v_b$ is the surface velocity, defined as:

\begin{equation} \label{eq3}
    v_b = \frac{\partial x}{\partial p}
\end{equation}
It describes how a point $x$ on the boundary changes in response to a perturbation of the parameter $p$. Figure 2(a) illustrates an 1D function with a discontinuous boundary condition. Despite the discontinuity at the boundary, the first-order derivative exists and follows the Leibniz integral rule:

\begin{equation} \label{eq4}
    \frac{d}{dp} \int_{a}^b f dx = \int_a^b\frac{\partial f}{\partial x}dx + f(x, b) \frac{d}{dp} b(p) - f(x, b) \frac{d}{dp} a(p)
\end{equation}

The first term in Eq. \ref{eq4} represents the direct differentiation of the integrand within the interior region, while the second and third terms account for the derivative with respect to the boundary location. The $d/ dp$ operator in the last two terms indicates how the boundary location changes in response to a perturbation of parameter $p$. In 2D scenarios, as shown in Figure 2(b), the Leibniz rule is generalized as Raynold’s transport theorem, as expressed in Eq. \ref{eq2} . In this context, the first term represents the derivatives within the interior region, and the second term corresponds to the differentiation with respect to the boundary. Eq. \ref{eq2}  assumes that the function values outside the region are zero. If the function values approaching the boundary from the interior and exterior are $f_+(x)$ and $f_-(x)$, respectively, Eq. \ref{eq2} can be written as:

\begin{equation} \label{eq5}
    \frac{d}{dp}L = \int_\Omega \frac{\partial}{\partial p} f(x, p)dV  + \int_{\partial \Omega} (v_b \cdot n) \Delta f dA
\end{equation}

\noindent where $\Delta f=f_- - f_+$.  Examples for gradient computation for segments and circles are provided in the supplementary information. In the following discussion, since our focus is on finding the derivative of shapes or the boundary of a design, we will primarily concentrate on the second term in Eq. \ref{eq5}.

\begin{figure}
  \label{fig:2}
  \
  \includegraphics[width=12cm]{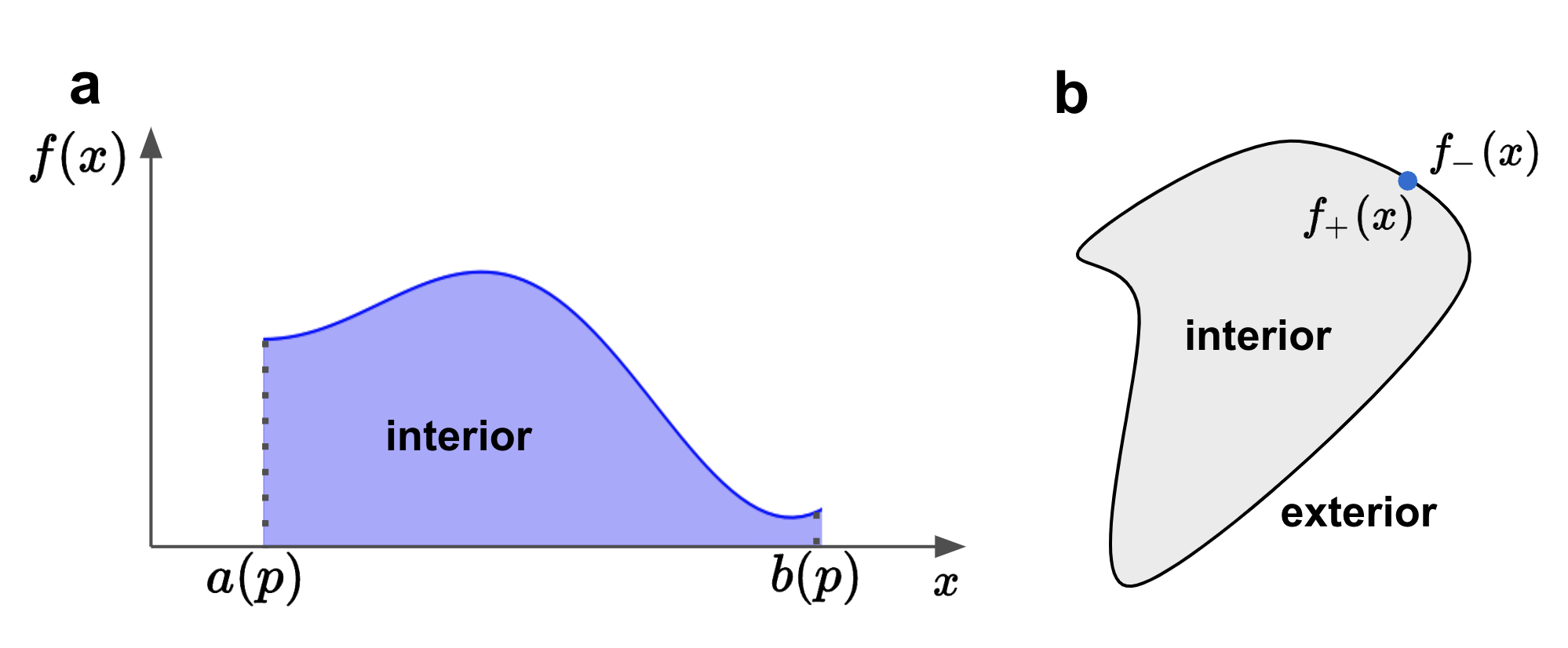}
  \caption{Illustration of shape differentiation: (a) Leibniz integral rule for 1D shapes and (b) Raynold’s transport theorem for 2D shapes. See the text for a detailed description.}
\end{figure}

\begin{figure}
  \label{fig:3}
  \
  \includegraphics[width=16cm]{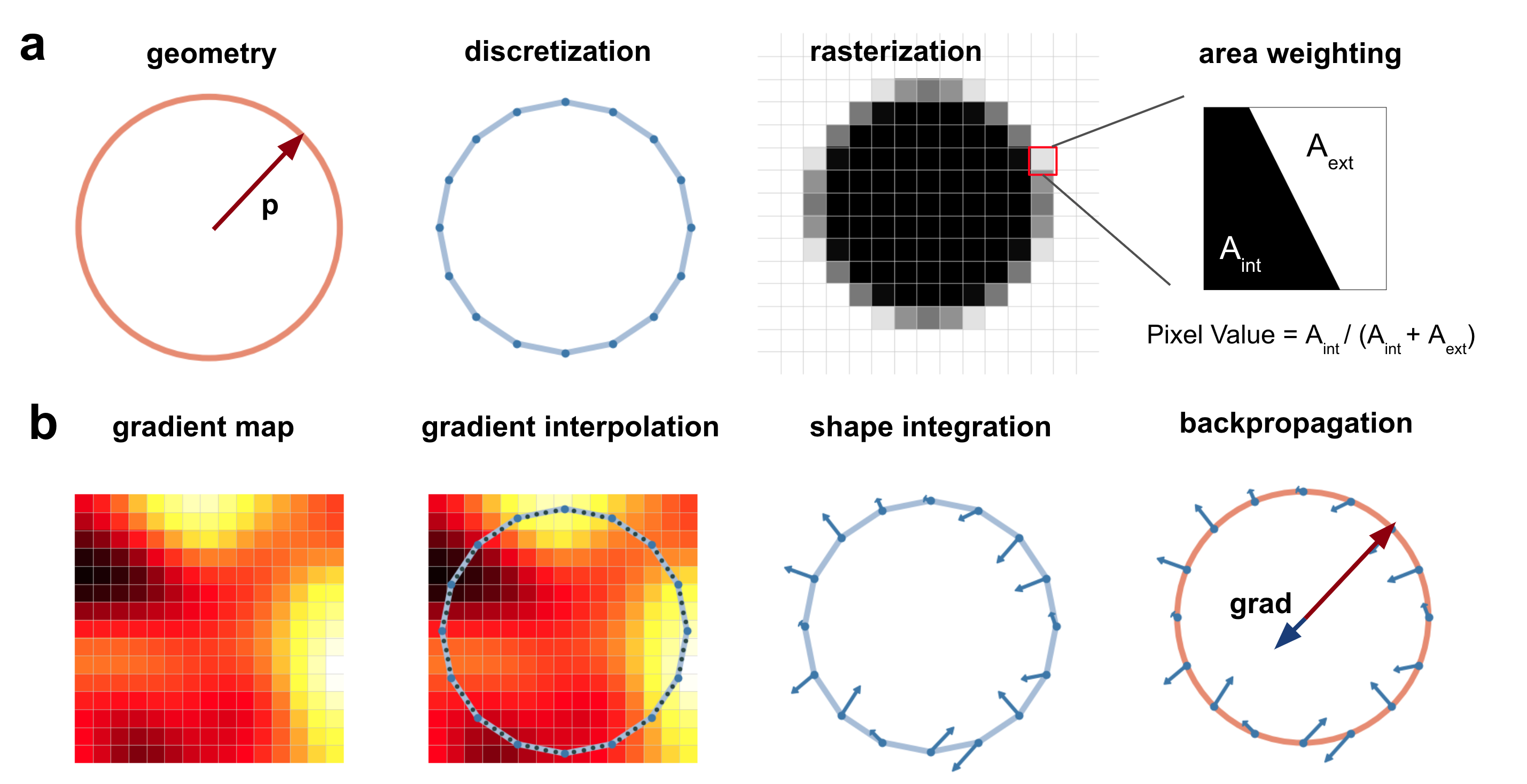}
  \caption{Workflow of differential shape optimization with explicit representation. (a) Forward structure generation: Given a geometry parameter $p$, the algorithm first generates its discretized representation using segments. With an image of a specific resolution, the algorithm rasterizes the discretized shape with subpixel smoothing. (b) Backward gradient calculation: Using the gradient map from a differentiable solver, we resample integration points on all segments of the discretized shape representation and interpolate the gradient values at these sample points. Raynold’s transport theorem is applied to determine the perturbation of each segment endpoint, and automatic differentiation is used to compute the gradient of the geometric parameter.
}
\end{figure}

\subsection{Algorithm Formulation }
Equipped with the shape derivative, we can now present the algorithm formulation for differential shape optimization using image representations. The workflow is summarized in Figure 3.

\subsubsection{Forward Rendering}
The forward rendering process rasterizes the input geometry into a binary image for simulation. This process is outlined in three steps, as shown in Figure 3(a):

\paragraph{\textbf{Geometry Construction}}   For a given shape parameter $p$, find its analytical parametric function. For example, a circle can be represented by:
\begin{equation} \label{eq6}
    x = r \cos \theta
\end{equation}
\begin{equation} \label{eq7}
    y = r \sin \theta
\end{equation}

where $\theta \in [0, 2\pi)$. Here we assume parameter $p$ is the radius $r$ of the circle. 

\paragraph{\textbf{Discretization}  }
Discretize the shape into segments. If the shape is represented by an explicit analytical function, this step automatically maintains gradients if the transformation is implemented using an automatic differentiation tool like PyTorch \cite{paszke2017automatic}. For instance, in the case of a circle as given in the equations, we can discretize $\theta$ to find the endpoints of segments $(x_i, y_i)$ that represent the discretized circle. For implicit representations, discretization algorithms often disrupt gradient calculations. We will address this issue in Section 4. 

\paragraph{\textbf{Rendering}} 
Rasterize the discretized shape onto an image. To maintain differentiability, the rasterized image must have a soft boundary. We use subpixel smoothing \cite{farjadpour2006improving} to achieve this, where the pixel value encompassing the boundary is weighted by the area of the shape's interior. This weighting method is crucial to ensure that gradients can be accurately computed during backward propagation.

\subsubsection{Backward Propagation}
Given the gradient map calculated from the differential solver, we can determine the gradients of the design parameter $p$ through the following steps:

\paragraph{\textbf{Resampling and Interpolation}}

Resample additional points along the segments discretized during forward rendering. Use an interpolation method to determine the gradient values at these resampled points.

\paragraph{\textbf{Gradient of Endpoints}}

Apply Eq. \ref{eq5} to compute the gradient of the endpoints of the discretized segments. Here, the endpoint coordinates $(x_i, y_i)$ are treated as design parameters. The analytical gradient of the segments is provided in the Supplementary Information.

\paragraph{\textbf{Calculation of Shape Gradients}}

Use Eq. \ref{eq6} to compute the gradients of the coordinates of the endpoints of all segments. The integral is evaluated over each segment, with $\Delta f$ representing the interpolated gradients at the sample points and $v_b$ as the surface velocity relative to the segment endpoints. The exact form of  $v_b$  for segments is detailed in the Supplementary Information.

\paragraph{\textbf{Backpropagation}}

With the gradients of the endpoints determined, use automatic differentiation to compute the gradient of the design parameter $p$. The gradient computation method for implicit representations is covered in Section 4.

The backward flow is shown in Figure 3(b). Note that we use Raynold’s transport theorem to compute the gradients of the segment coordinates $(x_i, y_i)$ and then derive the parameter gradients $\partial L/ \partial p$ using automatic differentiation. The theorem is not applied directly to derive the gradients of the design parameters $p$.

\subsubsection{Proof of Gradients
}
In the formulation of the backward propagation algorithm, we use $\Delta f=\partial f/ \partial x$, where $x$ represents the location of the sample points. However, Raynold’s theorem requires $f=f_--f_+$ as the integrand. In forward simulation, such quantities are not directly observable. Instead, we justify the use of gradients $\partial f /  \partial x$ in our scheme due to the soft boundary rendering.

To illustrate this, consider the Taylor expansion to perturb the function $f$ around the boundary:

\begin{equation} \label{eq8}
    f(x + \delta x)  \approx f(x) + \delta x \frac{\partial f}{\partial x}
\end{equation}
\begin{equation} \label{eq9}
    f(x - \delta x)  \approx f(x) - \delta x \frac{\partial f}{\partial x}
\end{equation}
The difference $\Delta f$ can be approximated as: 
\begin{equation} \label{eq10}
    \Delta f = f_- - f_+ = -2\delta x \frac{\partial f}{\partial x}
\end{equation}

Here, we introduce a scalar $2 \delta x$, which is related to the image discretization size. This scaling factor does not affect the direction of the gradients and can be ignored during optimization. Since our shape is discretized on a 2D canvas with a pixel area of $\Delta x^2$, the constant should be proportional to $\Delta x^2$ . However, it is important to note that the exact value of this scaling may differ slightly due to the residuals of the Taylor expansion. In later sections, we will demonstrate that this constant shift induced by discretization does not affect the smoothness of gradient computation during backpropagation.

\subsubsection{Extension to 3D}

The 3D gradients follow a similar approach to 2D, with a few modifications: In the forward process, the shape must be discretized into meshes. In the backward process, we sample on the mesh and compute the gradients of the vertices of all meshes. However, the complexity of handling 3D geometry and meshes introduces additional implementation challenges. To address these issues, we will primarily use implicit representation methods, which will be detailed in Section 4.

\subsection{\textbf{Gradient Validation}}
We use a disk example to validate the correctness and smoothness of the gradients derived from our scheme. In the forward process, we create a disk with radius $r$ assume the area of the disk as the loss function. In the backward process, we compute the gradients of area with respect to  $r$. Both forward and backward calculations have analytical forms. 

The area of the disk is:

\begin{equation} \label{eq11}
    A(r) = \pi r^2
\end{equation}
And the gradient of the area with respect to the radius is:
\begin{equation} \label{eq12}
    \frac{d}{dr} A(r) = 2 \pi r
\end{equation}

In the program, given the discretization size of the grid (or equivalently, the pixel size of the image), we compute the area as:
\begin{equation} \label{eq11}
    A(r; \Delta x) = \pi \Delta x^2 \sum_i I_i
\end{equation}
where $I_i \in [0, 1]$  represents the pixel values in the image.

Figure 4(a) plots the area with different radii using the forward algorithm with $\Delta x =0.1$, and compares it with Eq. (11). The area changes smoothly with respect to the radius due to subpixel smoothing. In Figure 4(b), we compare the gradient derived from our shape gradient algorithms with the analytical solution in Eq. \ref{eq12}. The gradient of the design parameter also varies smoothly with respect to the parameter $r$, demonstrating the smooth and accurate gradients required for gradient-based optimization.

To investigate the impact of discretization size $\Delta x$ on gradient computation, we plot the gradient of the radius for $r = 1$ with different $\Delta x$ values. The results are shown in Figure 4(c). As predicted by Eq. \ref{eq10}, the magnitude of the gradient is scaled by a constant factor proportional to $\Delta x^2$. Since a fixed $ \Delta x$ is chosen during optimization, this absolute scaling does not affect the optimization trajectory. In practical gradient computation, we can normalize the gradient by dividing it by $\Delta x^2$ to address the scaling issue caused by the resolution.

\begin{figure}
  \label{fig:4}
  \
  \includegraphics[width=16cm]{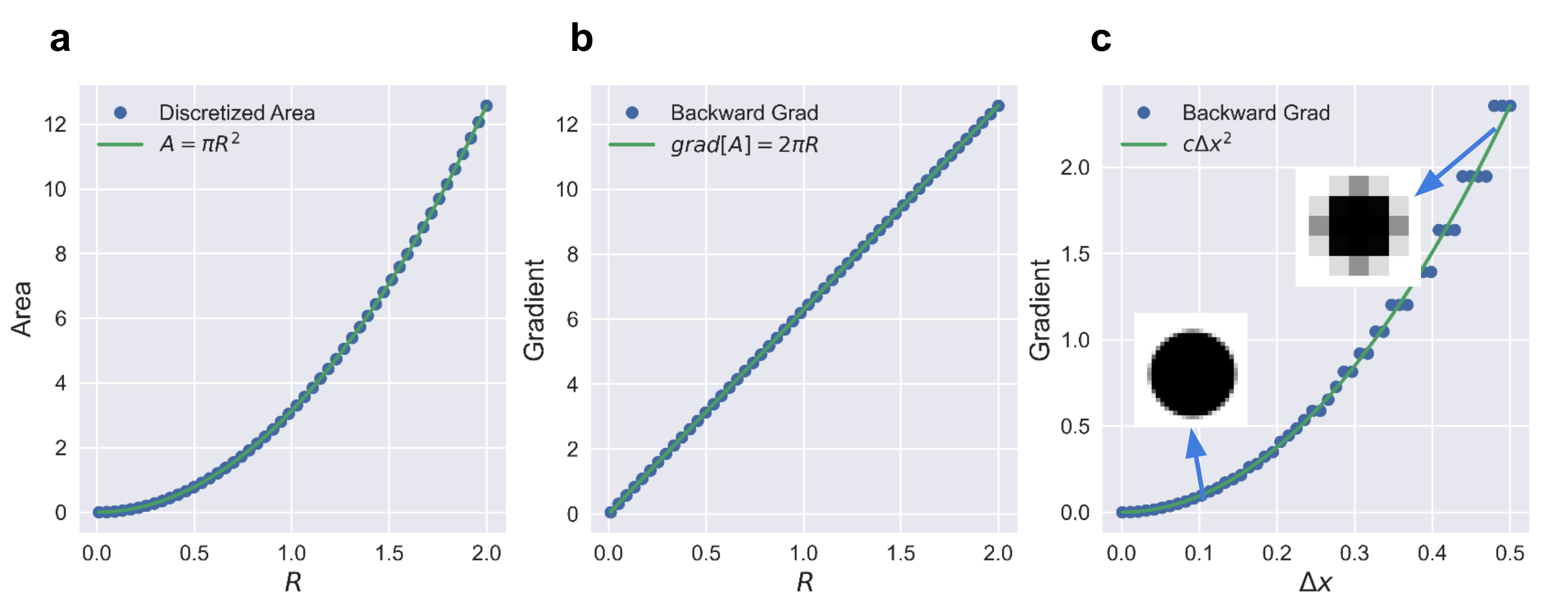}
  \caption{Validation of gradients for a disk. (a) The area of the forward rasterized disk with different radii, compared with analytical results for $\Delta x = 0.1$. (b) The gradient of the area with different radii, compared with analytical results for $\Delta x = 0.1$. (c) For a radius of $r=1.0$, the gradient is calculated with different $\Delta x$ values, compared with the analytical prediction.}
\end{figure}

\section{\textbf{Design Examples}
}To demonstrate that our algorithm is compatible with all pixel-represented simulation approaches, we applied it to three design examples using differential FDFD, FDTD, and RCWA. The software employed includes both open-source and commercial tools, with only the interfaces being revised to integrate our shape optimization method. The examples include a splitter in photonic integrated circuits, a diffractive lens at the micron scale, and a single-layered metasurfaces for beam steering. Our goal is not to achieve revolutionary performance but to evaluate the effectiveness of shape optimization. We emphasize the correctness of the gradients and the speed of convergence.

\subsection{\textbf{FDFD, Splitter}
}We first present the design of a splitter parametrized by polygons. We use Ceviche FDFD \cite{hughes2019forward} as the differential simulation to provide the gradients of the permittivities within the design region. The initial design and the simulated fields are illustrated in Figure 5(a). Light is injected from the left port and is equally split between the two right ports. The objective of the optimization is to maximize the total transmission power of the output ports, with the optimization wavelength fixed at 620 $nm$.

\begin{figure}
  \label{fig:5}
  \centering
  \includegraphics[width=12cm]{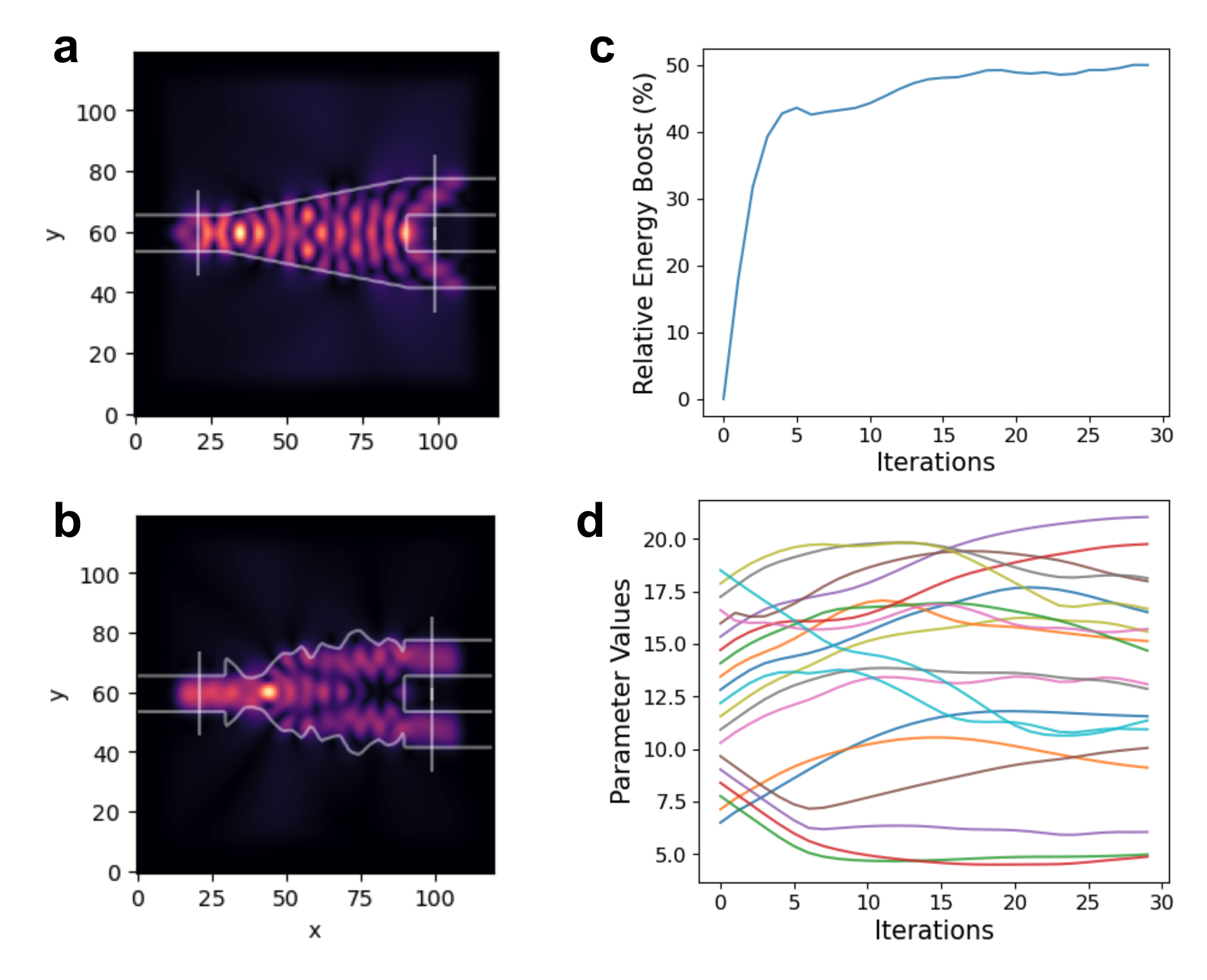}
  \caption{Design of a splitter with FDFD. (a) and (b) Design shapes before and after optimization. (c) Evolution of the relative figure of merit (FOM) improvement. (d) Evolution of all $y$-coordinates of the vertices of the design.}
\end{figure}

The design is parameterized by a polygon with a total of 40 vertices. Due to the symmetry of the design, we mirror-flipped the $y$-coordinates between the upper and lower edges. During optimization, we only adjust the $y$-positions of the vertices. The figure of merit (FOM) of the optimization is defined as maximizing the output electric fields of the two output ports:

\begin{equation}
    L = \int \|E_1(x)\| ^ 2 + \|E_2(x)\|^2 dx
\end{equation}
where $E_1$ and $E_1$ are the output fields at the two right ports. By maximizing this objective function, we aim to enhance the total transmission of the design and thereby reduce losses.

Since Ceviche FDTD, as a differential solver, naturally provides the gradient of all pixels in the design region, we can directly backpropagate the pixel gradients to the $y$-coordinates of the design using our shape optimization pipeline. Figure 5(b) shows the final design after optimization, and Figure 5(c) displays the improvement in transmission throughout the 30 iterations. The transmission continuously improves until it reaches its maximum value. The final optimized design exhibits non-intuitive shapes. Additionally, Figure 5(d) illustrates the variation of all design parameters.

\subsection{\textbf{FDTD, Diffractive Lenses }
}
The next example focuses on designing a compact diffractive lens to collimate a dipole source confined within a cylindrical cup coated with gold. The design configuration is shown in Figure 6(a). The cup has a height $d= 500 \: \mu m$ and a radius $r= 250 \: \mu m$. The dipole, oriented with $x$-polarization, is positioned at the center of the cylindrical cup. The objective is to use a diffractive lens set to collimate a dipole source with a wavelength of 620 $nm$ as effectively as possible in the center of the far field. 

\begin{figure}
  \label{fig:6}
  \centering
  \includegraphics[width=16cm]{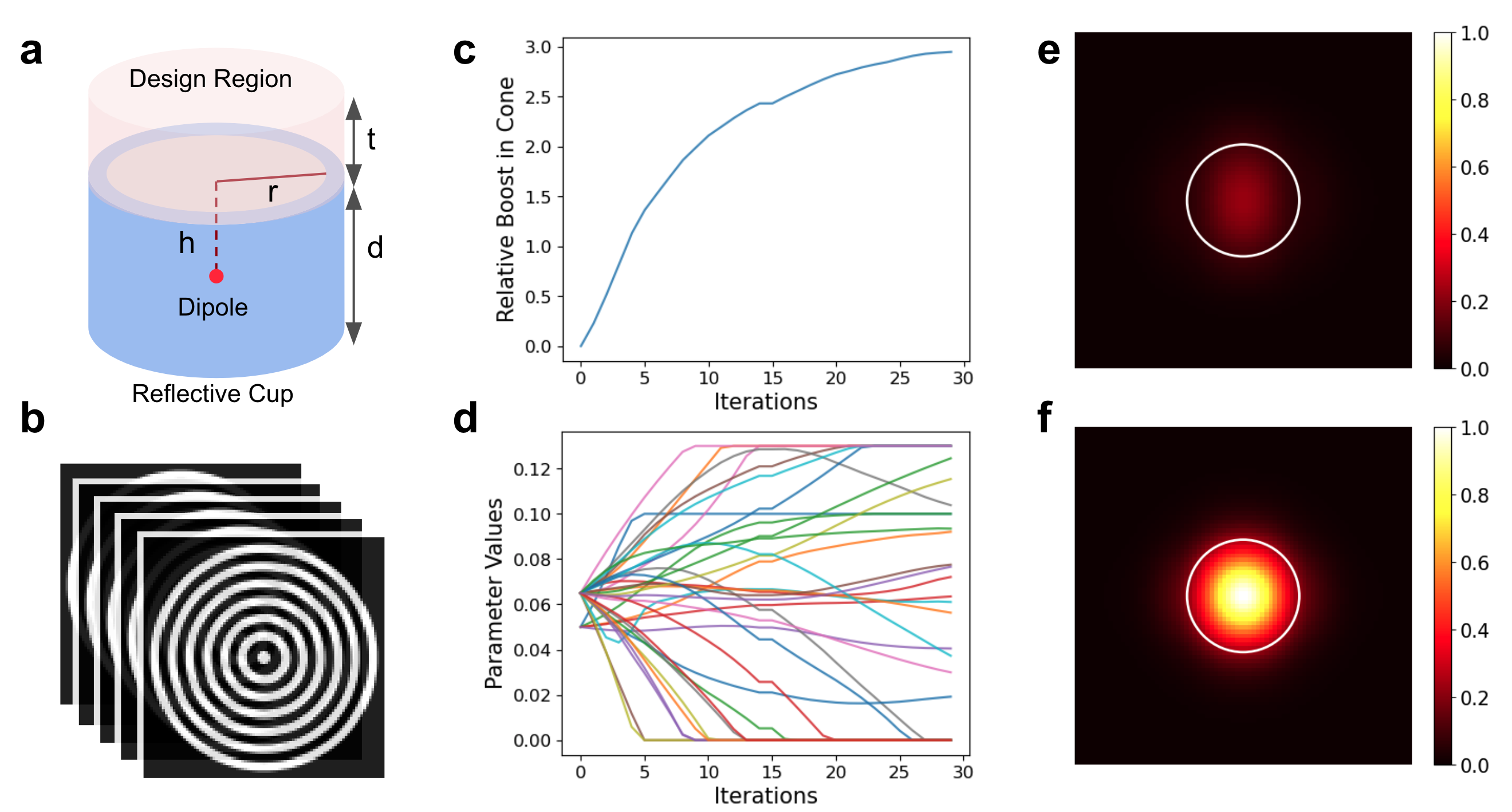}
  \caption{ Diffractive lens design with FDTD. (a) Simulation configuration: A dipole is positioned at the center of a cylindrical reflective cup with $d=500 \: nm$ and $r=250 \: nm$. The design region is placed on top of the cup. (b) Design region configuration: Five-layer cylindrical $\text{SiO}_2$ binary lenses, each consisting of 7 rings, are stacked to form the overall diffractive lens. (c) Evolution of the intensity in the far-field cone of interest. (d) Evolution of all ring radii throughout the optimization. (e) and (f) Far-field intensity before and after optimization, with circles indicating the region of interest.}
\end{figure}

The diffractive lens to be optimized is illustrated in Figure 6(b). The lens consists of 5 layers of ring structures, each 40 $nm$ thick, with a total design thickness of 200 $nm$. There are 7 rings in each layer, resulting in 35 design parameters for the entire diffractive lens. To create the lens designs, we first draw the circles defining the boundaries of the rings. For each ring, we create an outer disk with radius $r_{out} $ and a hollow disk with inner radius $r_{in}$. After generating 5 images of these rings, we extrude each image into a voxel to represent the 3D design and then stack them to form the voxel representation used in the simulation. 

We define the FOM as the sum of the far-field intensity within a 20-degree cone, where we aim for the lens to collimate the light as effectively as possible.

\begin{equation}
    L= \int_{\theta = 0^{\circ}}^{20^{\circ}} \|E(\theta) \|^2 d \theta 
\end{equation}

We use Lumerical 3D FDTD to simulate the design and apply an in-house adjoint method to derive the gradients of the density map. With the gradient map, we backpropagate the gradients to the design parameters using our framework. Figures 6(c) and 6(d) illustrate the relative improvement in far-field intensity within the cone and the evolution of all design parameters throughout the optimization, respectively. The FOM consistently improved, and the design converged in about 20 iterations. Note that constraints are applied to the design parameters to avoid invalid geometries. The final design's geometry is provided in the supplementary information. Figures 6(e) and 6(f) compare the far-field distribution before and after optimization, showing clear improvements with the light being more effectively collimated into the desired cone.

\subsection{\textbf{RCWA, Metasurface }
}
In the next example, we design a dielectric metasurface to deflect light by approximately 45 degrees at a wavelength of 980 $nm$. The optimization goal is to tune the parameters of the meta-atoms to maximize the diffractive efficiency at the desired angle. The configuration of the design is shown in Figure 7(a). The structure has a periodicity of $w=1.2 \mu m$ and a width of $h=0.3 \mu m$. Each period contains 8 elliptical meta-atoms with a height of 200 $nm$, spaced evenly across the structure. The definition of each meta-atom is illustrated in Figure 7(b), and the optimizable parameters are the long and short axes of all meta-atoms.

\begin{figure}
  \label{fig:7}
  \centering
  \includegraphics[width=16cm]{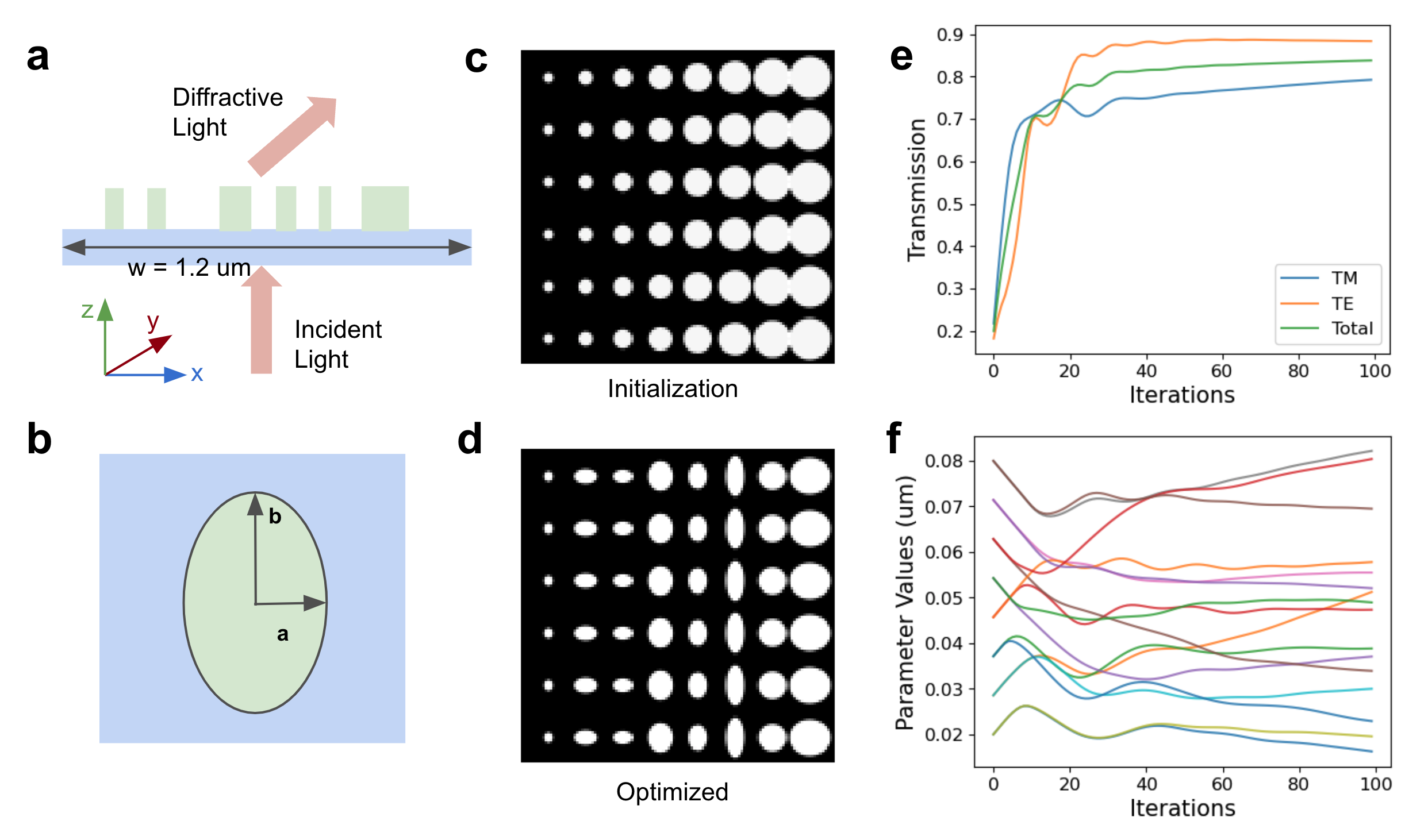}
  \caption{ Design of a metasurface with RCWA. (a) Side view of the metasurface: $\text{TiO}_2$ meta-atoms are coated on top of a glass substrate. The periodicity of the metasurface is $w = 1.2 \: \mu m$ in the $x$-direction and $h = 0.3 \: \mu m$ in the $y$-direction. (b) Each meta-atom is parameterized as an ellipse with long and short axes $a$ and $b$. (c) and (d) Design before and after optimization. (e) Diffractive efficiencies for TM and TE polarizations throughout the optimization, with the total diffractive efficiency also plotted. (f) Evolution of all design parameters.}
\end{figure}

The design starts with circles of linearly increasing diameters along the periodicity, as shown in Figure 7(c). The optimization objective is to enhance the first-order efficiency for both TE and TM polarization beams. This objective can be formulated as:

\begin{equation}
    L = \|E_{TE}\|^2 + \|E_{TM}\|^2
\end{equation}

\noindent where $E_{TE}$ and $E_{TM}$ are the electric fields of the first-order diffraction for TE and TM polarization, respectively.

We use an in-house differential RCWA to compute the gradients of the pixel density of the design and apply our shape optimization algorithm to backpropagate these gradients to the meta-atom diameters. The optimized design is shown in Figure 7(d). The variations in diffractive efficiency and the optimization trajectory are presented in Figures 7(e) and 7(f). The results demonstrate that the efficiency for both polarizations improves throughout the optimization process.

\section{\textbf{Implicit Shape Optimization}
}

\subsection{\textbf{Implicit Shape Representation}
}
In the preceding sections, we discussed the shape optimization pipeline using explicit representations. With explicit representations, the discretized vertex coordinates can be expressed as a function of the shape parameters, allowing the gradients of design parameters to be derived from the gradients of the vertices using Raynold’s transport theorem. However, more complex shapes, including 3D geometries, are often represented using implicit functions of the form:

\begin{equation} \label{eq17}
    f(x, p) = 0
\end{equation}
Implicit representation, sometimes referred to as level set representation depending on the context, poses challenges in discretization. Unlike explicit representations, where there is a direct mapping from design parameters to the discretized shape, implicit functions do not provide such straightforward mappings.

\subsection{\textbf{Gradient Algorithms for Implicit Representations}
}
To enable gradient calculation with implicit representations, several revisions are required to the algorithm presented in Section 2. These revisions include adjustments to the discretization scheme and the shape integration method.

\subsubsection{Discretization 
}
Discretization of a shape into polygons or meshes can be effectively handled using marching squares in 2D or marching cubes in 3D. In 2D scenarios, marching squares can extract the contour of a shape and represent it as a set of vertices and segments, which is ideal for rendering the shape onto an image. Using the extracted polygon contours, we apply the same rasterization algorithms described in Section 2 to render the shape on an image. For 3D shapes, we use marching cubes to obtain the mesh representations of the geometry and then rasterize the shape into a 3D image. The boundary voxel values are computed based on the ratio of the volume of the design interior to the total volume of the voxel.

\subsubsection{Shape Integration 
}
However, without explicit transformations from shape parameters, we cannot directly backpropagate the gradients from the vertices to the parameters. To address this issue, we begin by using the shape gradients for boundary integration that we derived earlier:

\begin{equation} \label{eq18}
    \frac{d}{dp}L =\int_{\partial \Omega} (v_b \cdot n)\frac{dL}{dz} dA
\end{equation}

where $z$ is a point on the boundary. We assume that the gradients on the boundary $dL/dz$ have been computed using interpolation methods. To proceed, we need to determine two variables: the surface velocity $v_b$ and the normal direction of the boundary $n$. The normal direction can be computed using the equation:

\begin{equation} \label{eq19}
    n = \frac{\nabla f}{\| \nabla f\|}
\end{equation}

where $\nabla f$ represents the gradient of the implicit function with respect to spatial coordinates. By plugging Eq. \ref{eq19} and the expression of $v_b$ in Eq. \ref{eq3} into Eq. \ref{eq18}, and after rearranging the orders of these terms, we get:

\begin{equation} \label{eq20}
    \frac{d}{dp}L =\int_{\partial \Omega}  \nabla f \frac{\partial x}{\partial p} \frac{1}{\|\nabla f \|}\frac{dL}{dz} dA
\end{equation}

Using chain rule, we note that: 

\begin{equation} \label{eq21}
    \nabla f \frac{\partial x}{\partial p} = \frac{\partial f}{\partial p}
\end{equation}

Replacing Eq. \ref{eq21} into Eq. \ref{eq20}, we finally arrive at: 

\begin{equation} \label{eq22}
    \frac{d}{dp}L =\int_{\partial \Omega}  \frac{\partial f}{\partial p} \frac{1}{\|\nabla f \|}\frac{dL}{dz} dA
\end{equation}

This equation provides the backward gradient integration. The discretization of the equation is:

\begin{equation} \label{eq23}
    \frac{d}{dp} L \approx \sum_i \frac{\partial f(z_i, p)}{\partial p} \frac{1}{\|\nabla f(z_i, p) \|} \frac{dL}{dz} \Delta A_i
\end{equation}

\noindent where $\Delta A_i$ represents the size of the segment or mesh. This equation indicates the procedure of backward propagation for implicit shapes. As with the explicit representation, we need to resample the segments or meshes and determine the values of  for all resampled points. With these values, we can regenerate the function value $f(z_i,p)$ on all resampled points and perform backpropagation, assuming the gradient to be backpropagate on each point $z_i$ is 

\begin{equation} \label{eq24}
    g_i = \frac{dL}{dz} \frac{1}{\| \nabla f(z_i, p) \|} \Delta A_i 
\end{equation}
 
Notice that $\nabla f$ can be easily retrieved using automatic differentiation. An example demonstrating the equivalence between the formulations for explicit and implicit representations is included in the Supplementary Information.

\subsection{\textbf{Application of Implicit Shape Optimization}
}
Since any explicit shape representation can be expressed implicitly, we can use implicit shape optimization as a general framework for the optimization pipeline. The advantage of implicit representation lies in its ability to encompass a much larger degree of freedom within the design space, where explicit representations may fall short. The forward-backward algorithm described above can be directly applied to implicit representations, including bicubic interpolation \cite{vercruysse2019analytical}, Fourier series \cite{liu2020topological}, and even neural implicit representations \cite{liu2020compounding}. Our approach ensures a consistently binarized design space while allowing seamless gradient propagation to the design parameters.

\section{\textbf{Conclusion}}
In summary, we have introduced a general framework for differentiating shapes represented in binary images with respect to their parameters, which can be applied in various density-based optical simulations, including FDTD, FDFD, and RCWA. Our algorithm effectively rasterizes the density map of optical structures with subpixel smoothing and accurately derives gradients, regardless of the image resolution. We have detailed methods for both explicit and implicit shape representations, covering a wide range of possibilities, including simple geometries, level sets, and neural representations, applicable in modern photonic and optical design. The efficacy and generalizability of our algorithm are demonstrated through three design cases using different differential solvers. Our framework provides foundational building blocks for general differential optimizations in optical and photonic design. 






\newpage
\bibliographystyle{unsrt}  
\bibliography{references}  

\newpage
\section*{Supplementary Information}
\label{sec:supp}

\subsection*{S1. Shape Integration for Segments}

The core of the shape gradient integration involves integrating along the discretized segment, which is parametrized by two vertices. For a segment defined by the endpoints  $(x_0, y_0 )$ and $(x_1, y_1)$. Our goal is to find the analytical gradient of $v_n$ in Eq.5 for the two endpoints.  These gradients can then be propagated back to the shape parameters. The point $(x, y)$ on the segment can be represented by the parameterization equation:

\begin{equation} \label{eqs1}
    x = (1 - t) x_0 + t x_1
\end{equation}

\begin{equation} \label{eqs2}
    y = (1 - t) y_0 + t y_1
\end{equation}

\noindent where $t \in [0, 1]$. We can find all derivatives of $(x, y)$  with respect to the endpoint variables as follows:

\begin{align}
\frac{\partial x}{\partial x_0} &= 1 - t & \frac{\partial x}{\partial y_0} &= 0   & \frac{\partial x}{\partial x_1} &= t  & \frac{\partial x}{\partial y_1} &= 0 \nonumber \\
\frac{\partial y}{\partial x_0} &= 0 & \frac{\partial y}{\partial y_0} &= 1 - t   & \frac{\partial y}{\partial x_1} &= 0  & \frac{\partial y}{\partial y_1} &= t
\label{eq:eqs3}
\end{align}

The surface velocity of the point $(x, y)$ thus can be found using the definition Eq. \ref{eq3}:
\begin{align}
\frac{\partial}{\partial x_0}(x,y) &= [1 - t,  0]^T & \frac{\partial}{\partial y_0}(x,y) &= [0, 1-t]^T  & \frac{\partial}{\partial x_1}(x,y) &= [t, 0]^T  & \frac{\partial}{\partial y_1}(x,y) &= [t, 0]^T 
\label{eqs4}
\end{align}

There are two normal directions of a segment with two opposite directions, and we choose
\begin{equation} \label{eqs5}
    n = \frac{1}{l}[-(y_1 - y_0), x_1 - x_0]^T 
\end{equation}

\noindent in our gradient computation, where $l$ is the length of the segment. We thus can derive $v\cdot n$ for all endpoints coordinates: 

\begin{align} 
\frac{\partial (x,y)}{\partial x_0} \cdot n &= \frac{1}{l} (1 - t) (y_1 - y_0) & \frac{\partial (x,y)}{\partial y_0} &= \frac{1}{l} (1 - t) (x_1 - x_0) \nonumber \\
\frac{\partial (x,y)}{\partial x_1} \cdot n &= -\frac{t}{l} (y_1 - y_0) & \frac{\partial (x,y)}{\partial y_1} &= \frac{t}{l} (x_1 - x_0) 
\label{eqs6}
\end{align}

We further can solve $t$ from Eq. \ref{eqs1} and Eq. \ref{eqs2}:
\begin{equation} \label{eqs7}
    t = \frac{x - x_0}{x_1 - x_0} = \frac{y - y_0}{y_1 - y_0}
\end{equation}

We replace $t$ in Eq. \ref{eqs6} and finally achieved the $v \cdot n$ for all variables:

\begin{align} 
\frac{\partial (x,y)}{\partial x_0} \cdot n &= \frac{1}{l} (y - y_1) & \frac{\partial (x,y)}{\partial y_0} &= - \frac{1}{l}  (x - x_1) \nonumber \\
\frac{\partial (x,y)}{\partial x_1} \cdot n &= -\frac{1}{l} (y - y_0) & \frac{\partial (x,y)}{\partial y_1} &= \frac{1}{l} (x - x_0) 
\label{eqs8}
\end{align}

In the above derivation, we assume the length of the segment is $l$. When plugging in the $v \cdot n$ into the shape derivative, we can treat $l$ as the discretization lengths $dl$ of the integration, and thus $1/dl$ is canceled with the $dl$ in Eq. \ref{eq5}. 

\subsection*{S2. Example for Explicit and Implicit Shape Gradients }

Here we derive the shape integral for a circle parametrized by radius $r$ and its center $(x_0, y_0)$, to illustrate the equivalence of our algorithm for explicit and implicit representation. 

\subsubsection*{S2.1 Explicit Representation}
A circle parametrized by $r$ and $(x_0, y_0)$ can be formulated by the equation:
\begin{equation} \label{eqs9}
    (x - x_0)^2 + (y - y_0)^2 = r^2
\end{equation}

The function can be explicitly represented using the parameter equation:
\begin{equation} \label{eqs10}
    x = x_0 + r \cos \theta 
\end{equation}
\begin{equation} \label{eqs11}
    y = y_0 + r \sin \theta 
\end{equation}

\noindent where $\theta \in [0, 2\pi) $. By definition Eq. \ref{eq3}, the surface velocity of a point $(x, y)$ on the circle with respect to all variables can be calculated as:

\begin{align}
\frac{\partial}{\partial r}(x,y) &= [\cos \theta,  \sin \theta]^T & \frac{\partial}{\partial x_0}(x,y) &= [1, 0]^T  & \frac{\partial}{\partial y_0}(x,y) &= [0, 1]^T 
\label{eqs12}
\end{align}

The normal vector of the circle can be easily derived as:
\begin{equation}\label{eqs13}
    n = [\cos \theta, \sin \theta]^T
\end{equation}

The term  $v \cdot n$ for all variables are:
\begin{align}
\frac{\partial (x,y) }{\partial r} \cdot n &= 1  & \frac{\partial (x,y) }{\partial x_0} \cdot n &= \cos \theta  & \frac{\partial (x,y) }{\partial y_0} \cdot n &= \sin \theta
\label{eqs14}
\end{align}

We can replace $\theta$ using Eq. \ref{eqs10} and Eq. \ref{eqs11}:

\begin{align}
\frac{\partial (x,y) }{\partial r} \cdot n &= 1  & \frac{\partial (x,y) }{\partial x_0} \cdot n &= \frac{x - x_0}{r}  & \frac{\partial (x,y) }{\partial y_0} \cdot n &= \frac{y - y_0}{r}
\label{eqs15}
\end{align}

Finally, we can put everything together into the shape integral Eq. \ref{eq5}:
\begin{align}
\frac{d}{dr}L &= \int_{\partial \Omega} \frac{\partial L}{\partial z} dl  & \frac{d}{dx_0}L &=  \int_{\partial \Omega} \frac{x - x_0}{r}\frac{\partial L}{\partial z} dl  & \frac{d}{dy_0}L &=  \int_{\partial \Omega} \frac{y - y_0}{r}\frac{\partial L}{\partial z} dl
\label{eqs16}
\end{align}

\subsubsection*{S2.2 Implicit Representation}
As comparison, we can directly start from the implicit representation of the circle:
\begin{equation} \label{eqs17}
    f(x, y; p) = r^2 - (x - x_0)^2 - (y - y_0)^2
\end{equation}

where $p = [r, x_0, y_0]$. In Eq. \ref{eq22}, We need to derive $\partial f /\partial p$ and $1 / \|\nabla f\||$ respectively of the implicit function. We can find that: 

\begin{equation} \label{eqs18}
    \nabla f = [-2(x - x_0), -2(y - y_0)]^T
\end{equation}

And thus,

\begin{equation} \label{eqs19}
    \|\nabla f\| = 2 \sqrt{(x - x_0)^2 + (y - y_0)^2} = 2r
\end{equation}

For each parameter, the term $\partial f /\partial p$ can be found as:
\begin{align}
\frac{\partial f }{\partial r} &= 2r  & \frac{\partial f}{\partial x_0}&= -2(x - x_0)  & \frac{\partial f}{\partial y_0} &= -2(y - y_0)
\label{eqs20}
\end{align}

Plugging all terms into Eq. \ref{eq22}, we find the derivative for all variables as:
\begin{align}
\frac{d}{dr}L &= \int_{\partial \Omega} \frac{\partial L}{\partial z} dl  & \frac{d}{dx_0}L &=  \int_{\partial \Omega} \frac{x - x_0}{r}\frac{\partial L}{\partial z} dl  & \frac{d}{dy_0}L &=  \int_{\partial \Omega} \frac{y - y_0}{r}\frac{\partial L}{\partial z} dl
\label{eqs21}
\end{align}

This shows that the shape integral of explicit and implicit are identical. 

\subsection*{S3. Optimized Diffractive Lens Shapes}
\begin{figure}
  \label{fig:s1}
  \includegraphics[width=18cm]{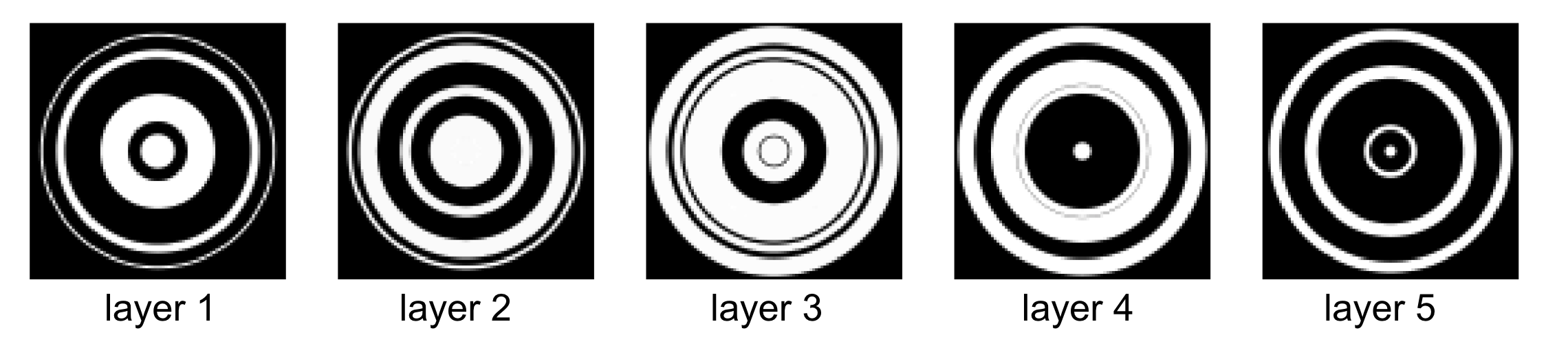}
    \captionsetup{labelformat=empty}
  \caption{Figure S1: The optimized shape of the diffractive lens in Section 3.2 from the bottom layer to the top.}
\end{figure}

\end{document}